\documentclass[%
 aip,
 amsmath,amssymb,
 reprint,%
]{revtex4-1}

\usepackage{graphicx}% Include figure files
\usepackage{dcolumn}% Align table columns on decimal point
\usepackage{bm}% bold math
\usepackage[utf8]{inputenc}
\usepackage[T1]{fontenc}
\usepackage{mathptmx}
\usepackage{etoolbox}

\usepackage{hyperref}
\usepackage[capitalize]{cleveref}
\hypersetup{
    colorlinks,
    citecolor=blue,    filecolor=blue,
    linkcolor=blue,    urlcolor=blue
}

\makeatletter
\def\@email#1#2{%
 \endgroup
 \patchcmd{\titleblock@produce}
  {\frontmatter@RRAPformat}
  {\frontmatter@RRAPformat{\produce@RRAP{*#1\href{mailto:#2}{#2}}}\frontmatter@RRAPformat}
  {}{}
}%
\makeatother
\begin{document}

\title{Decoupling acceleration and wiggling in a laser-produced Betatron source}
% Force line breaks with \\

%\preprint{}

\author{Julien Gautier}
\affiliation{Laboratoire d’Optique Appliquée, ENSTA Paris, CNRS, Ecole polytechnique, Institut Polytechnique de Paris, 828 Bd des Maréchaux, 91762 Palaiseau, France}

\author{Igor A Andriyash}
\affiliation{Laboratoire d’Optique Appliquée, ENSTA Paris, CNRS, Ecole polytechnique, Institut Polytechnique de Paris, 828 Bd des Maréchaux, 91762 Palaiseau, France}

\author{Andreas D\"opp}
\affiliation{Ludwig-Maximilians-Universität München, Am Coulombwall 1, 85748 Garching, Germany}

\author{Michaela Kozlova}
\affiliation{Institute of Plasma Physics, CAS, Za Slovankou 3, 182 21 Prague 8, Czech Republic}

\author{Aimé Matheron}
\affiliation{Laboratoire d’Optique Appliquée, ENSTA Paris, CNRS, Ecole polytechnique, Institut Polytechnique de Paris, 828 Bd des Maréchaux, 91762 Palaiseau, France}

\author{Benoit Mahieu}
\affiliation{CEA, DAM, DIF, F-91297 Arpajon, France}

\author{Cedric Thaury}
\affiliation{Laboratoire d’Optique Appliquée, ENSTA Paris, CNRS, Ecole polytechnique, Institut Polytechnique de Paris, 828 Bd des Maréchaux, 91762 Palaiseau, France}

\author{Ronan Lahaye}
\affiliation{Laboratoire d’Optique Appliquée, ENSTA Paris, CNRS, Ecole polytechnique, Institut Polytechnique de Paris, 828 Bd des Maréchaux, 91762 Palaiseau, France}

\author{Jean-Philippe Goddet}
\affiliation{Laboratoire d’Optique Appliquée, ENSTA Paris, CNRS, Ecole polytechnique, Institut Polytechnique de Paris, 828 Bd des Maréchaux, 91762 Palaiseau, France}

\author{Amar Tafzi}
\affiliation{Laboratoire d’Optique Appliquée, ENSTA Paris, CNRS, Ecole polytechnique, Institut Polytechnique de Paris, 828 Bd des Maréchaux, 91762 Palaiseau, France}

\author{Pascal Rousseau}
\affiliation{Laboratoire d’Optique Appliquée, ENSTA Paris, CNRS, Ecole polytechnique, Institut Polytechnique de Paris, 828 Bd des Maréchaux, 91762 Palaiseau, France}

\author{Stéphane Sebban}
\affiliation{Laboratoire d’Optique Appliquée, ENSTA Paris, CNRS, Ecole polytechnique, Institut Polytechnique de Paris, 828 Bd des Maréchaux, 91762 Palaiseau, France}

\author{Antoine Rousse}
\affiliation{Laboratoire d’Optique Appliquée, ENSTA Paris, CNRS, Ecole polytechnique, Institut Polytechnique de Paris, 828 Bd des Maréchaux, 91762 Palaiseau, France}

\author{Kim Ta Phuoc}
\email{kim.ta-phuoc@cnrs.fr}
\affiliation{Centre Lasers Intenses et Applications, Université de Bordeaux, CNRS, CEA, UMR 5107, 33400, Talence, France}

\date{\today}

\begin{abstract}

Betatron radiation is produced in Laser Plasma Accelerators when the electrons are accelerated and simultaneously wiggle across the propagation axis \cite{Rousse2004}. The mechanisms of electron acceleration and X-ray radiation production follow different scaling laws \cite{Corde2013}, and the brightest X-ray radiation is often produced for an electron beam with a lower quality in terms of energy and divergence. Here, we report a laser-driven Betatron X-ray source where the plasma density profile is tailored in order to separate the acceleration and wiggler stages, which allows for the independent optimizations of acceleration and X-ray production. We demonstrate this concept experimentally, and show that the Betatron photon energy can be controlled by adjusting the length of the plasma wiggler. This scheme offers a path to overcome the limitations of conventional Betatron sources, enabling the production of bright, stable, energetic, and collimated X-ray beams.

\end{abstract}

\maketitle

In laser-plasma acceleration, an intense femtosecond laser pulse propagates through an underdense plasma and drives a highly nonlinear wave in the form of a series of co-propagating ion cavities \cite{Pukhov2002}. Electrons trapped in one of these cavities are accelerated to the relativistic energies and at the same time oscillate transversely across the propagation axis. This oscillating motion of relativistic electrons produces the synchrotron-like X-rays commonly known as the Betatron radiation \cite{Kiselev2004, Rousse2004, Schnell2015}. The characteristics of this emission depend exclusively on the electron orbits \cite{Corde2013}, and here we explore a way to enable the decoupling of acceleration and wiggling in a Betatron source.

The electron orbit in the ion cavity is roughly sinusoidal, and is characterized by the longitudinal oscillation period $\lambda_\beta$, the transverse oscillation amplitude $r_\beta$, and the Lorentz factor of electrons $\gamma_e$. The $r_\beta$ is related to the size of the electron beam, and is defined by the injection process, while the oscillation period is determined by $\gamma_e$ and the electron plasma density $n_{pe}$. The characteristic energy and number of emitted photons are given respectively by $E_c [\text{eV}] = 5.24 \times 10^{-21} \gamma_e^2 n_{pe} [\text{cm}^{-3}]r_\beta[\text{$\mu$m}]$ and $N = 3.31 \times 10^{-2}K$, where $K$ is the wiggler strength parameter $K=1.33 \times 10^{-10} \sqrt{\gamma_e \; n_{pe}[\text{cm}^3]}r_\beta[\text{$\mu$m}]$. These expressions show that energy and flux of the source can be improved by increasing $K$ through the increase of $\gamma_e$, $n_{pe}$ or $r_\beta$ \cite{NatPhys2010Kneip, TaPhuoc2008, FerriPRAB2018, Yu2018, Kozlova2020, Xiao2020, Tomkus2020, Rakowski2022, Mishra2022, Hojbota2023}. However, in the typical situation where electron acceleration and radiation emission occur simultaneously, the parameters $\gamma_e$ and $n_{pe}$ cannot be optimized independently. As it turns out, the increase in $n_{pe}$, which is beneficial for X-ray production, results in a faster acceleration dephasing and leads to the lower $\gamma_e$. Furthermore, the combination of several electron injection mechanisms, that occur at higher plasma densities, gives rise to the shot-to-shot fluctuations \cite{Corde2013Injection, Doepp2017}. To avoid the trade-off between the quality of electron beam and number of X-ray photons, one may decouple the acceleration and wiggling processes. Such decoupling was discussed theoretically in various configurations \cite{TaPhuoc2008,FerriPhysRevLett2018,Zhu2020}, and here we explore this concept experimentally aiming to increase the radiation energy and flux, while maintaining or even improving the source collimation. 

%Consequently, for all Betatron X-ray sources developed so far, the brightest X-ray radiation originates from a degraded electron beam. 

\begin{figure}
 \includegraphics[width=1.\linewidth]{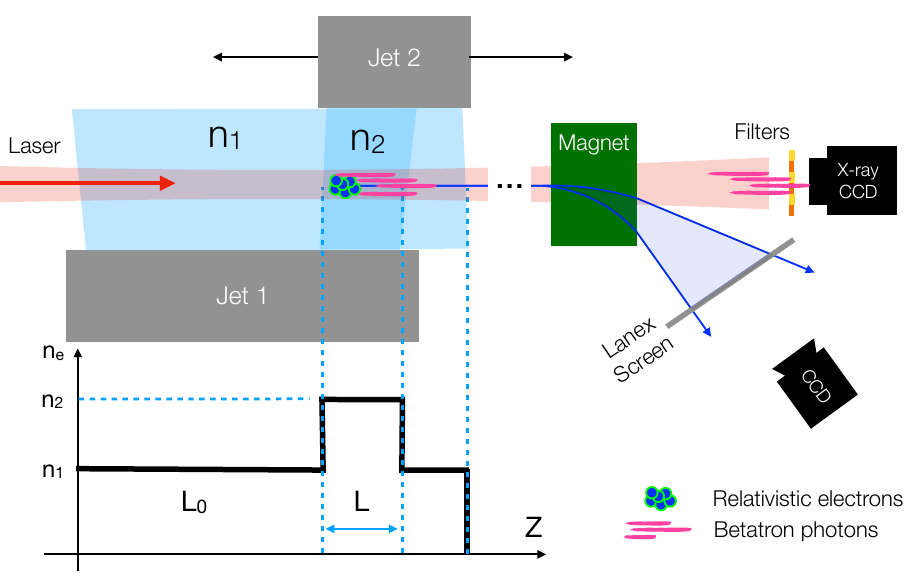}%
\caption{Schematic representation of the experimental setup. Electrons are accelerated by LWFA in the first gas jet which density $n_1$ is optimized to produce the highest electron beam quality in terms of energy and divergence. The second gas jet, with a density $n_2 = 2 \times n_1$ is used as a plasma wiggler to optimize the production of X-rays from the electron beam. The length $L$ of the plasma wiggler can be tuned by moving the second jet.}\label{Fig1}%
\end{figure}

The principle of the considered scheme is illustrated in Figure~\ref{Fig1}. It relies on the plasma density profile tailored along the laser propagation, which is equivalent to tailoring the wakefield itself \cite{TaPhuoc2008,FerriPRAB2018,Yu2018,Kozlova2020,Tomkus2020,Gustafsson2024}. In this configuration, the laser first propagates through a gas jet with a density $n_1$, which is optimized to efficiently drive a laser plasma accelerator (LPA) over a length $L_0$. A second jet is then placed on top of the first jet, creating a density step with density $n_2$ and length $L$, as shown in Figure~\ref{Fig1}. This step is used as a plasma wiggler and the density here is set to be approximately twice that of the LPA stage. This setup allows control over the length of the plasma wiggler, but the shadowgraph diagnostic revealed a shock at the interface between the two jets. Nonetheless, no correlation was observed between the shock and the behaviors of electrons and X-rays beams presented below.

%Indeed the radius of the cavity is given by $r_b(z) = 2\sqrt{a_0}/k_p(z)$ where $k_p(z) = 2\pi / \omega_p(z)$ with $\omega_p(z) = (4\pi e^2 n_e(z)/m)^{1/2}$ the plasma frequency and $e$, $m$ the electron charge and mass. The radial and longitudinal electric fields in the cavity are respectively given by $E_r(z) = E_0(k_p(z)^2 r/2)$ and $E_z = E_0(k_p(z) z +r_b(z) - v_g(z) t)/2$ where $E_0 = k_p(z) mc^2/e$, $r$ is the distance to the cavity axis and $v_g(z)$ the laser group velocity. 

Let us consider and discuss the phenomena that occur as the electron and laser beams enter and propagate in the plasma wiggler. When laser passes from the region of low plasma density to the one with a higher density, the ion cavity shrinks. As this happens, the electrons located closer to the cavity front remain within the same cavity \cite{TaPhuoc2008}, and their phase in the wakefield resets to the stronger acceleration, so they gain more energy through this so-called re-phasing \cite{DoeppPoP2016}. The parameter $K$ can also increase due to the higher plasma density, thus leading to a higher X-ray flux and energy, but without a significant change in divergence. Alternatively, for the electrons located further from the ion cavity front, the ion cavity shrinking causes particles to slip into the next period of the plasma wave. These electrons are exposed to the consecutive de-focusing and de-phasing wakefields that force some particles to exit the wakefield, while others gain higher divergence and energy spread, but can still be re-trapped in the cavity and oscillate with the increased transverse amplitude. This results in the higher flux and energy of the X-rays, accompanied by a significant increase in divergence. 

Furthermore, depending on the laser Rayleigh length and the plasma density $n_1$, the laser pulse may get significantly depleted and diffracted during the propagation, and not longer be able to efficiently drive wakefield in the plasma wiggler section. In such cases, the electron beam itself may drive a wakefield \cite{Hidding2019, Kurz2021}. Similar situation occurs when the injected electron charge becomes large (e.g. a case of ionization injection), and over-loads the wake, thus becoming the dominant driver. Electrons then would continue oscillating and producing Betatron radiation without de-phasing, but with the growing energy spread. Depending on $L_0$, $L$, $n_1$ and $n_2$, either of the discussed scenarios or their combination may take place and affect the Betatron radiation. Moreover, these individual contributions may be very challenging to distinguish in the experiment.

The experiment was conducted at Laboratoire d’Optique Appliquée (LOA) using the "Salle Jaune" laser system (2~Joules / 30~femtoseconds). The laser was focused with a spherical mirror used off-axis with an f-number of 18. Residual aberrations were corrected using a deformable mirror. The focal spot diameter was approximately 25~$\mu$m (FWHM). The first gas jet consisted of a 1~mm~$\times$~5~mm slit operating with a mixture of 99\% helium and 1\% nitrogen. 

 \begin{figure}
 \includegraphics[width=0.8\linewidth]{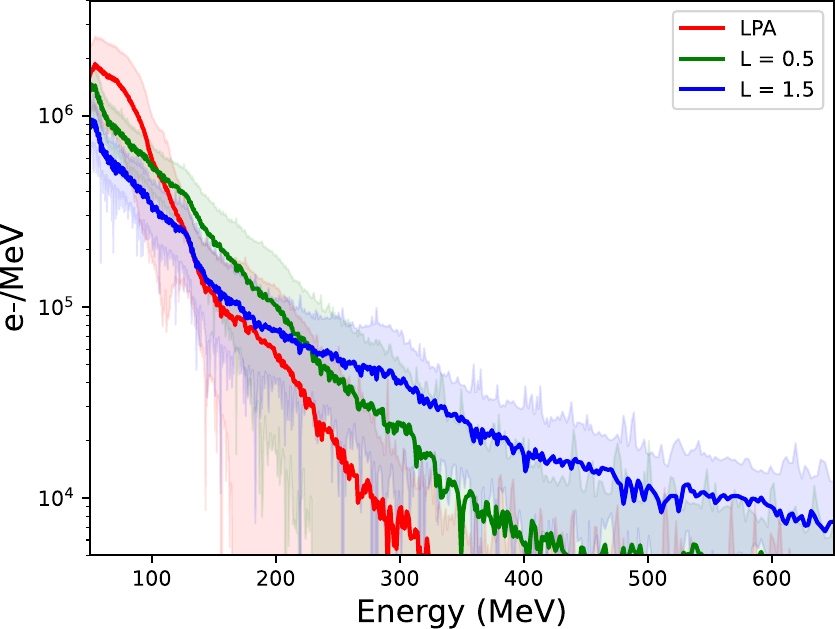}%
 \caption{Average electron spectra (over 50 shots) at the output of the LPA without plasma wiggler (red) and when the length of the plasma wiggler is $L = 0.5$ mm (green) and $1$ mm (blue).}%
 \label{Fig2}
 \end{figure}

Betatron radiation produced in the LPA was measured using an X-ray CCD (indirect detection QuadRO - Princeton Instruments). The laser light was blocked by a 15~$\mu$m aluminum filter. A 100~$\mu$m Mylar window provided air-vacuum separation, and a 500 $\mu$m beryllium filter was placed in front of the camera. The system was optimized for detection in the spectral range from $\sim$5 to a few tens of keV. The camera was placed 70~cm from the source so that the size of the X-ray beam fits the CCD chip ($\sim$~4.5~cm~$\times$~4.5~cm). This setup enabled the measurement of the X-ray beam profile and therefore the integrated flux. The spectrum was estimated by measuring the X-ray signal through metallic filters. A 5~cm~$\times$~5~cm matrix of 16~$\times$~4 filters was used, consisting of 10~$\mu$m, 30~$\mu$m, and 105~$\mu$m copper and 10 $\mu$m aluminum filters placed in front of the camera. The signal from the identical filters was interpolated to reconstruct the spatial profile of the X-ray beam. A synchrotron spectrum with a critical energy, $E_c$, was assumed, and a numerical image of the X-ray beam passing through the filters was constructed for comparison with the experimental image. The best value of $E_c$ was then determined by minimizing the error between the experimental and numerical images. This analysis method does not allow for precise spectrum reconstruction, nonetheless it gives a good approximation of a Betatron spectrum and it provides a critical energy with a resolution, of about 1.5 keV, that is sufficient for the purpose of this experiment.

\begin{figure}
 \includegraphics[width=0.8\linewidth]{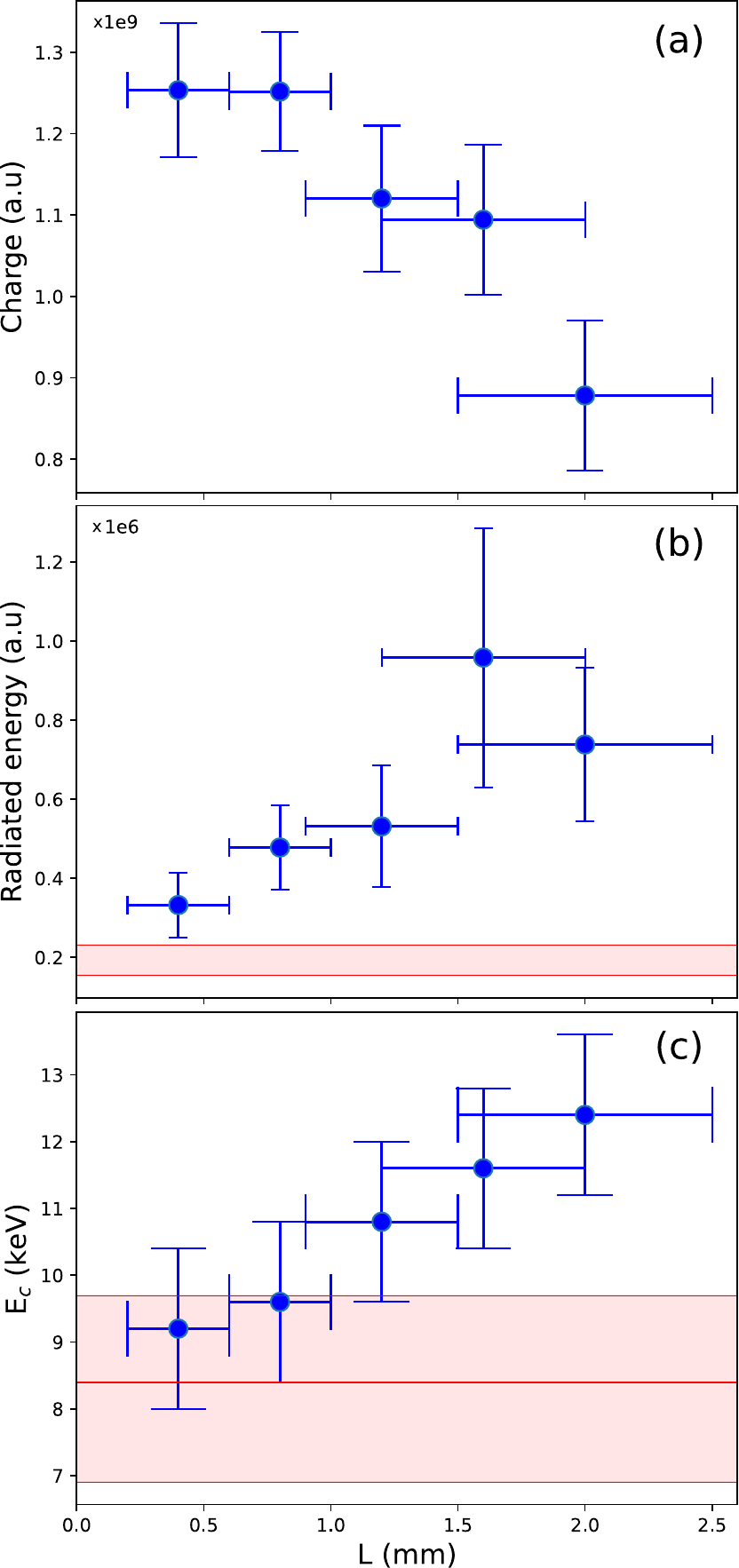}%
 \caption{(a) Electron beam charge as a function of $L$. The charge is estimated using the electron spectrometer diagnostic. (b) Radiated X-ray energy measured as a function of $L$. (c) Critical energy of the X-ray radiation as a function of $L$. The mean critical energy in the LPA stage and error bar are represented in red. Each point in (a,b and c) corresponds to the average over 50 shots.}%
 \label{Fig3}
 \end{figure}

In the experiment, we have first optimized the LPA stage without the second jet in order to maximize the electron energy, and the optimum was obtained for the plasma electron density, $n_{pe} = 8 \times 10^{18}$~cm$^{-3}$. In our configuration, electrons were primarily injected into the wakefield by ionization injection \cite{Pak2010}. The beam with a typical broad, thermal spectrum extending up to $\sim$200~MeV, and the divergence of approximately 5~mrad (FWHM) was stably obtained with the total charge that ranged between 50~pC and 100~pC above 50~MeV. The X-ray beam produced in the LPA stage had the divergence of ${\sigma_\theta^\text{xray} = 32 \pm 2}$~mrad (FWHM) and the critical energy of $E_c = 8.5 \pm 1.5$~keV.

% ###### AJOUT

The second jet was then placed at the output of the LPA. This distance is close to the dephasing length in jet 1, which is $\sim$ 5 mm in our parameter regime. The density in the plasma wiggler was set to $n_2 \sim 2 \times n_1$. Addition of the plasma wiggler has affected the total acceleration by simultaneously reducing the $L_0$ and adding the density step, as shown in Figure~\ref{Fig1}. In Figure~\ref{Fig2} we plot the electron spectra obtained at the output of the LPA stage and with the plasma wiggler at two positions, $L = 0.5$~mm and $L = 1.5$~mm. One may see, that the characteristic electron energy increases with $L$, which we attribute to re-phasing. \cite{Katsouleas1986, Guillaume2015}. %Indeed, the size $r_b$ of the cavity is reduced by $\sqrt{n_2/n_1}$ in the plasma wiggler. The electron slips back into an accelerating phase of the wake, at a distance $r_b (1-\sqrt{n_1/n_2})$ to the middle of the cavity and can be further accelerated. The energy gain expected in the second jet depends on the driver (laser beam or electron beam). This will be discussed later.

The ion cavity contraction may also cause some electrons to exit the wakefield entirely, resulting in a loss of charge. Figure~\ref{Fig3}(a) shows the integrated charge $Q$ as a function of $L$. For $L<1$~mm, the charge remains comparable to that measured with the first jet only. For a longer wiggler, $L>1$~mm, we see the decrease of the charge, which can be explained by the electron de-phasing and laser depletion that leads to deceleration and loss of the effective charge. 
%Finally, the plasma wiggler has also affected the electron beam divergence. We observe an increase by about 5 times.

In the wiggler section, the plasma wave can be driven either by the laser (LWFA) and/or by the electron beam (PWFA), depending on the length $L$ and the charge of the bunch. While the used diagnostics do not allow us to distinguish between these two wakefield regimes directly, there are a few considerations that suggest the transition from LWFA to PWFA in our experiment. First, the length of the LPA ranged from $\sim 3$~mm to 5~mm, while the laser depletion length estimates around $\sim 5$~mm, which means that in the wiggler section a significant fraction of laser energy may already be lost. Second, in the high-density wiggler, the laser slows down and this typically leads to the dephasing or significant charge loss, which was not observed in the experiment for a broad range of $L$. Alternatively, an active laser driver could also trigger injection, thus increasing the charge of lower energy fraction of the beam, and this was not observed neither. From these considerations, we may assume that in the plasma wiggler the wakefield is rather driven by the electron bunch than by the laser.

The produced X-ray emission includes the radiation produced in the LPA and in the plasma wiggler, and we may study both contributions by varying the wiggler length. In Figure~\ref{Fig3}(b), we plot the total measured energy of emitted X-rays as a function of $L$. This value is obtained by spatially integrating the X-ray beam profile recorded by the camera. One may see, that the radiated energy first increases with $L$, reaching its maximum at approximately $L \sim 1.6 \, \text{mm}$, and then it starts to drop with $L$. 

The linear increase of X-ray signal follows the growing number of betatron oscillations performed by the particles, their energy and wiggling frequency. The nonlinear LPA dynamics may also produce growth of electrons beam emittance, which results in the growing $K$ and thus enhancing the X-ray emission. The drop of the signal after, $L > 1.5 \, \text{mm}$, correlates with the drop of the effective charge observed in Figure~\ref{Fig3}(a). This interplay results in the existence of an optimal $L$ for the maximum X-ray flux.

The X-ray spectrum is characterized by the critical photon energy. It is represented as a function of $L$ in Figure~\ref{Fig3}(c). For the shorter acceleration length ${L\lesssim 1}$~mm, we observe the critical photon energies similar to those measured in the LPA stage. It then grows linearly with the wiggler length reaching $E_c = 12 \pm 1.5$~keV for $L=2$~mm.

 \begin{figure}
 \includegraphics[width=1.\linewidth]{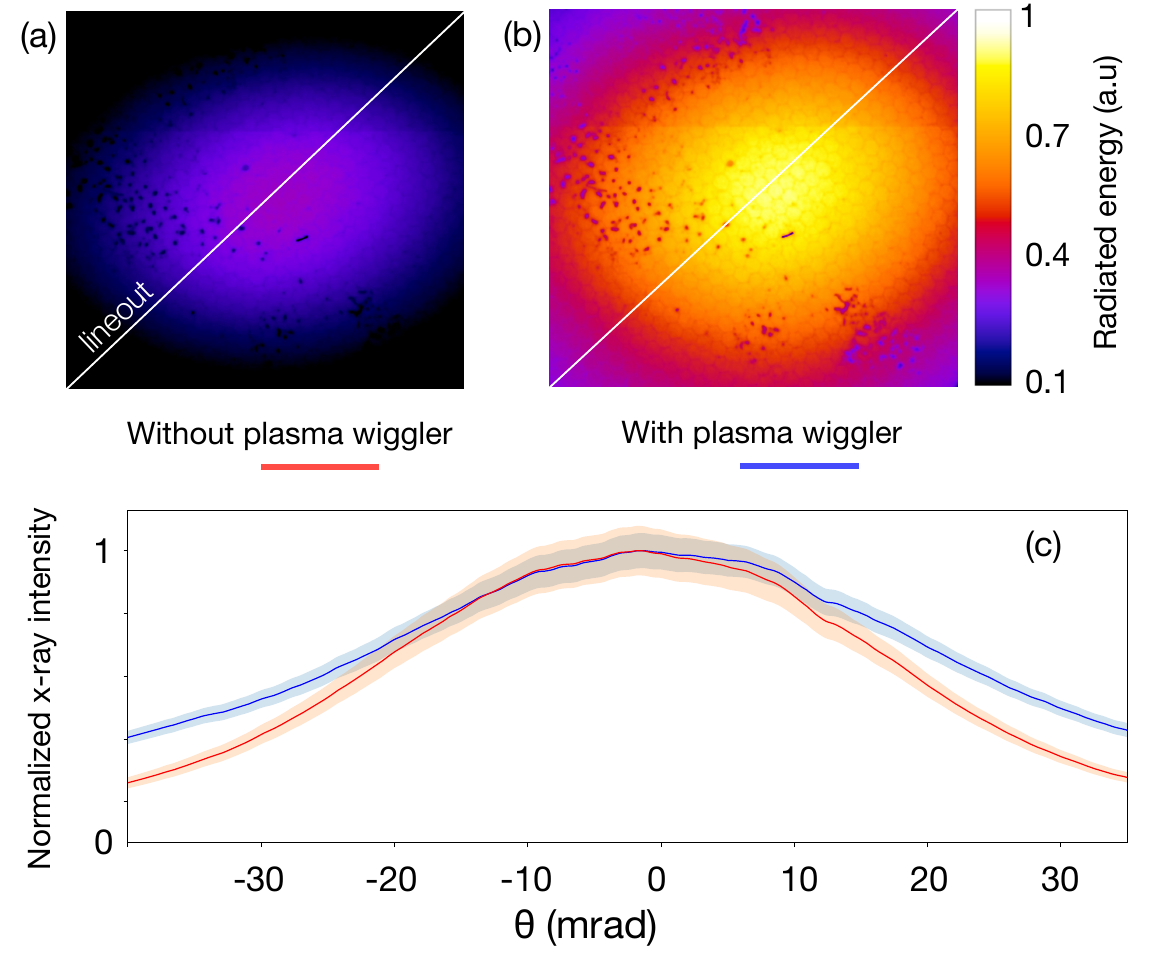}%
 \caption{X-ray beam profiles averaged over 20 shots without the plasma wiggler (a) and with the plasma wiggler (b). (c) Lineouts of the x-ray beam profiles along the white lines in (a) and (b).}%
 \label{Fig4}
 \end{figure}

The spatial profiles of X-rays beams obtained in the experiment, shown in Figure~\ref{Fig4}, indicate the radiation angular divergence of ${\sigma_\theta^\text{xray} = 39 \pm 2}$~mrad (FWHM), which was found to be roughly independent of $L$. The betatron source divergence naturally includes contributions from the angular divergence of electron beam, $\sigma_\theta^\text{e-}$, and from the intrinsic synchrotron radiation divergence ${\sigma_\theta^\text{SR} \approx K / \gamma_e}$. In the LPA betatron sources, the latter term typically dominates, and it scales with plasma and electron parameters as $\propto \sqrt{n_{pe} / \gamma_e}$. In the plasma wiggler both $n_{pe}$ and $\gamma_e$ increase, and the resulting change of X-rays divergence turns out to be insignificant. This highlights one of the key advantages of our decoupling scheme: by tuning only second stage, we can optimize the X-ray production while maintaining consistent beam collimation.

%######## Discussion Div Electrons / vs div Betatron

%The divergence of the x-ray beam does not evolves the same as the divergence of the electron beam. This indicates that the electron beam divergence continues to evolve after the electrons have stopped emitting a significant amount of X-ray radiation, likely in the residual gas after the density step. where beam hosing instability can develop \cite{Whittum1990}.

%######## Discussion Stabilité

Finally, decoupling should, in principle, lead to improved stability, as the electron beam itself is stable. However, in our experiment, no improvement in the X-ray beam stability was observed. We believe this can be attributed to shot-to-shot variations in the interface between the two gas jets, and therefore variations of $L$, as evidenced by the shadowgraphy images.

%##############################

\begin{figure}
 \includegraphics[width=0.8\linewidth]{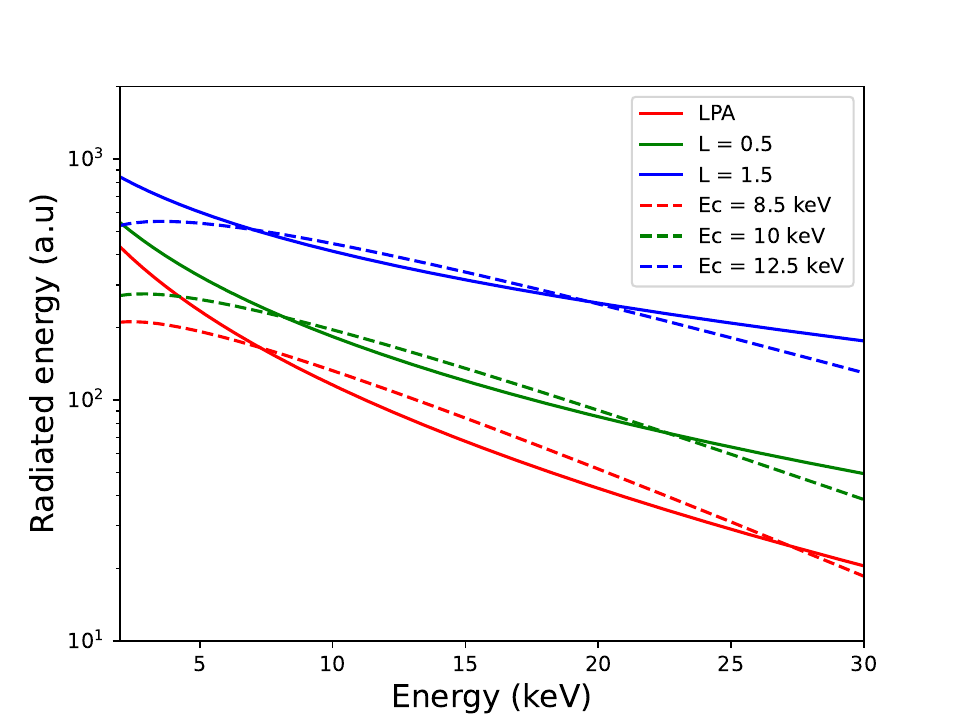}%
 \caption{Solid lines represent the X-ray spectra calculated using the electron spectra from Figure~\ref{Fig2} assuming a transverse oscillation amplitude of $r_\beta = 1$ micron and a number of oscillations in the LPA that 6x and 2x that in the plasma wiggler respectively for $L=0.5$~mm and $L=1.5$~mm. Dotted red, green and blue lines represent synchrotron functions that can fit these calculated spectra.}%
 \label{Fig5}
\end{figure}

In order to interpret these experimental results, we use a test-particle model. We calculate the X-ray emission considering the measured electron spectra (see Figure~\ref{Fig2}). For simplicity, we assume electron spectra in the form of the distribution ${N = \exp(-E/E_f)}$, where characteristic electron temperatures that fit the measurements are found $E_f = 50$~MeV, $E_f = 75$~MeV and $E_f = 100$~MeV, corresponding respectively to the LPA section alone and to both LPA and plasma wiggler sections with $L=0.5$~mm and $L=1.5$~mm. In our modeling we consider sinusoidal electron trajectories with the amplitude and number of oscillation $r_\beta$ and $N_\beta$ being the free parameters in both the LPA and plasma wiggler. The radiation is calculated using the classical formula for synchrotron radiation \cite{jackson_classical_1999}. The best agreement with experimental data was achieved for $r_\beta = 1$~$\mu$m (typical for Betatron sources \cite{PRE2006Shah, PRL2006TaPhuoc}), and when the numbers of oscillations in the LPA are respectively 6x and 2x that in the plasma wiggler respectively for $L=0.5$~mm and $L=1.5$~mm. The latter ratio is consistent with the ratio of the useful lengths of the LPA and plasma wiggler. In Figure~\ref{Fig5} we plot the simulated spectra along with the synchrotron function fits for $E_c = 8.5$~keV, $E_c = 10$~keV and $E_c = 12.5$~keV for the configurations without the plasma wiggler and with $L=0.5$~mm and $L=1.5$~mm, respectively. Note, that although the single synchrotron function may not fit the entire energy range we achieve a reasonable agreement over an energy range from 5~keV to 30~keV, which corresponds to the spectral region where our X-ray spectrometer is most efficient. Finally, the integration of the spectra (i.e. the total radiated energy) is in good agreement with Figure 3b.

In conclusion, the results presented above demonstrate the possibility of decoupling acceleration and wiggling, enabling the production of Betatron X-rays from the optimized electron beams. The enhancement of the X-ray radiation is attributed to a combination of re-phasing and an increase in $K$ within the plasma wiggler. We did not observe an increase in the X-ray beam divergence, ruling out a significant increase in $r_\beta$ as the origin of the increased X-ray flux and energy.

While this scheme does not allow for a significant increase of the Betatron radiation in our parameter range, it may become very attractive at higher, Petawatt laser powers. Indeed, achieving electron energies in the GeV range requires dephasing lengths on the centimeter scale and plasma densities in the range of  $10^{17}$ $\text{cm}^{-3}$. This would result in a Betatron wavelength on the order of several millimeters, which is not favorable for X-ray production. Instead, the GeV electrons propagating in a plasma wiggler with the densities around of the order of $10^{20} \, \text{cm}^{-3}$ will generate the high-brightness Betatron radiation in the few-hundred-keV energy range.

\bibliography{bibliography.bib}

%\noindent LaTeX formats citations and references automatically using the bibliography records in your .bib file, which you can edit via the project menu. Use the cite command for an inline citation, e.g.  \cite{Figueredo:2009dg}.

\section*{Acknowledgements}

This work was supported by the French ANR agency under Grant FemTraXS No. ANR-21-CE30-0013 and grant LABX Campus France 50048WB.

\end{document}